# Novel High-Scalability Architecture for Photonic Deep Learning

Yuxin Sun[1], Chun Gao[1], Jin Xie[1], Pan Wang[1], Zejie Yu[1], Yiwei Xie[1], Huan Li[1*], Daoxin Dai[1*]

[1]State Key Laboratory for Modern Optical Instrumentation, Center for Optical & Electromagnetic Research, College of Optical Science and Engineering, International Research Center for Advanced Photonics, Zhejiang University, Zijingang Campus, Hangzhou 310058, China

*e-mail: lihuan20@zju.edu.cn, dxdai@zju.edu.cn

**Abstract**

Photonic computing promises ultrafast and energy-efficient artificial intelligence. However, existing photonic neural networks (PNNs) remain functionally shallow and difficult to scale. Here we establish a theory-guided framework showing that power stability and complex-field correlation are the fundamental prerequisites for scalable, coherent PNNs. Building on these macroscopic principles, we introduce the Coherent, Compensated and Cross-connected (C3) unit — an architecture that integrates coherent nonlinearity, active loss compensation and native optical residual connectivity. Implemented on a silicon-on-insulator platform, the C3 unit provides reconfigurable activation functions and dynamic energy stabilization without external amplification. We validate this framework using a width-constrained spiral benchmark, in which the C3 unit substantially improves parameter utilization and power robustness relative to incoherent nonlinearities. In a high-complexity 1,623-class recognition task, our C3-enabled coherent residual network (CoP-ResNet) achieves a top-1 accuracy of 77.92%, whereas non-residual architectures fail to converge. Together, these results offer a physically grounded, theory-guided pathway toward greater optical processing depth, laying the foundation for next-generation, large-scale photonic computing architectures.

**Introduction**

Artificial intelligence (AI), particularly deep learning, has achieved remarkable performance across diverse applications[1–3], driving exponential growth in computational demand[4]. In the post-Moore era, electronic architectures encounter fundamental limitations in power efficiency and computational density as transistor scaling nears atomic-scale boundaries[5,6], demanding alternative computing paradigms. Photonic computing offers exceptional potential for real-time AI inference owing to its ultralow latency and high energy efficiency[7,8]. In particular, silicon photonics (SiPh) has demonstrated state-of-the-art performance in optical linear processing[9–18], delivering record-breaking energy efficiency[19] and low latency[20].

This momentum has shifted attention from isolated linear processors to photonic neural networks (PNNs)[21–27] in pursuit of greater expressivity and model scale. One common approach is to single-layer photonic accelerators through high-speed electronic interfaces[25,26]. While this approach enables substantial model depth and parameter scaling, each layer incurs electronic control and interconnect overhead, progressively accumulating latency, power consumption and bandwidth constraints. An alternative pathway is to maximize optical processing depth[24,27], thereby reducing electronic overhead. However, existing demonstrations remain functionally shallow, typically comprising no more than four layers and approximately $10^4$ reconfigurable parameters—several orders of magnitude fewer than those of modern AI models[3,28,29]. This gap persists due to three interrelated challenges. First, a theoretical foundation for photonic computing is largely absent, preventing principled scaling, most designs mimic electronic neurons rather than exploiting the complex-valued nature of light. Second, physical constraints—specifically cumulative attenuation and inherently weak optical nonlinearities—compromise layer-to-layer stability and model expressivity. Third, current PNNs are architecturally constrained, largely restricted to multilayer perceptron (MLP)- or convolutional neural network (CNN)-style feedforward forms, precluding more advanced structures such as residual pathways[28,30].

Here, we present a framework that addresses these challenges. As illustrated in Fig. 1a, the propagation of high-dimensional random optical fields gives rise to emergent statistical behavior[31,32]. As layer width increases, pre-activation fields converge toward well-defined statistical distributions, and we identify two macroscopic order parameters—the average optical power $q$ and the normalized complex-field correlation $c$—as the key descriptors governing scalable network behavior. Crucially, our analysis reveals that only coherent amplitude activation functions (AFs) can guide the network dynamics into the expressive, scalable regime shown in Fig. 1b. By contrast, incoherent intensity-based nonlinearities invariably confine the system to a non-separable state, while purely linear processing (coherent or incoherent) invariably leads to unstable power dynamics. For lossy systems, the theory further shows that residual pathways are necessary to maintain stable access to this expressive regime.

Guided by these macroscopic dynamics, we introduce the Coherent, Compensated, and Cross-connected (C3) unit, which implements coherent amplitude nonlinearity, power stabilization, and residual connectivity to enable scalable coherent photonic networks (Fig. 1c). On a silicon-on-insulator platform, the C3 unit provides tunable coherent AFs and effective energy compensation. To validate scalability, we construct two coherent PNNs using C3 units: a five-class spiral classification and a 1,623-class Omniglot[33] recognition task. In the spiral task, C3-based networks achieve superior accuracy and robustness, particularly in width-constrained settings, while maintaining stable performance over a broad input-power range. For the more challenging Omniglot task, we demonstrate CoP-ResNet, a coherent photonic residual architecture enabled by the C3 unit. This architecture attains 77.92% test accuracy—surpassing optical networks lacking residuals (< 16%) and electronic baselines using complex-valued GELU (76.88%). These results establish a principled pathway from statistical field theory to device-level implementation and system-level scalable architectures.

The resulting framework provides a physically grounded basis for constructing deeper, more expressive photonic networks, laying a foundation for future optical implementations of advanced architectures.

**Results**

**Mean-field Dynamics and Scaling Principles.** The scalability of coherent PNNs is governed by the propagation dynamics of complex optical fields (Supplementary Equation S1). We extend mean-field and phase-transition theory to complex-valued photonic architectures. As layer width increases, pre-activation fields (**h**$^l$) converge toward CSCG distributions, independent of the specific nonlinearity (Supplementary Fig. S1). This statistical self-averaging reduces the microscopic dynamics to the evolution of two macroscopic order parameters at layer $l$: the mean optical power $q^l$ (Supplementary Equation S7) and the normalized complex-field correlation $c^l$ (Supplementary Equation S13) between any two inputs. Their evolution is governed by iterative maps:

$$q^l = \mathcal{Q}(q^{l-1}|\sigma_w, \sigma_b)$$

$$c^l = \mathcal{C}(c^{l-1}, q^*|\sigma_w, \sigma_b)$$

(1)

The fixed point $q^*$ of $\mathcal{Q}$-map determines whether optical power attenuates, amplifies, or propagates stably. The fixed point $c^*$ of $\mathcal{C}$-map characterizes the similarity between two input signals as processed by the network. The stability of the trivial solution $c^*=1$ (identical inputs) is governed by the Lyapunov exponent $\chi_1$ (Supplementary Equation S20), which dictates the network's operational phase. In the order phase ($\chi_1 < 1$), correlations contract toward 1, eliminating discriminability. Conversely, in the chaotic phase ($\chi_1 > 1$), the $c^*=1$ becomes unstable and a new stable fixed point $c^*<1$ emerges, enabling stable feature separation. When $\chi_1 \gg 1$, correlations approach 0, causing extreme noise sensitivity. The transition at $\chi_1 = 1$ defines an edge-of-chaos regime. Full derivations are provided in Supplementary Section 2.

Applying this framework reveals the specific requirements for scalable coherent PNNs. Fig.2a shows that linear amplitude responses (both coherent and incoherent) lacks a no non-zero stable fixed point under zero bias ($\sigma_b = 0$), causing to power decay or explosion. An amplitude AF is required to govern the non-zero stable power ($q^* > 0$). Correlation dynamics (Fig.2b) reveal a second condition: incoherent amplitude AFs inevitably collapse correlations to $c_* = 1$ (ordered phase), coherent amplitude nonlinearities can sustain a non-trivial $c_* < 1$, enabling expressive propagation. The phase diagram of $\chi_1$ (Fig.2c) further highlight this distinction. Incoherent designs remain strictly ordered ($\chi_1 < 1$). Coherent AFs exhibit a clear phase boundary separating ordered and chaotic regimes. However, in practical photonic systems, losses constrain the effective weight variance ($\sigma_w \leqslant 0.5$ for square layers, Supplementary Equation S23), pushing networks into the ordered phase. This reveals the fundamental bottleneck: the linear transformations become contraction mappings—operators with singular values < 1—due to inevitable anisotropic loss. Under repeated application, these contractions progressively compress the high-dimensional signal manifold, driving dynamics toward a fixed point where expressive capacity vanishes. Crucially, compensating for optical power loss alone cannot reverse this collapse of information geometry, as it merely rescales the signal without recovering the lost orthogonal components within the high-dimensional state space. Therefore, scalable architectures require mechanisms that actively preserve correlation structure against contraction. Residual connections exemplify such a mechanism, providing a compensatory pathway that bypasses contractive operations and stabilizes the informational dynamics.

Together, these results show that scalable coherent photonic deep neural networks (CoP-DNNs) require a phase-preserving coherent amplitude AF combined with cross-layer connectivity. Detailed numerical simulation methods and additional results are provided in Supplementary Section 3.

**Device Design and Operational Principle.** A wide range of photonic nonlinearities have been explored, broadly classified as all-optical (AO)[34–36] and synergistic photoelectronic (SPE)[21,37–41]. AO approaches rely on intrinsic material nonlinearities and preserve full optical coherence, but typically require high intensities or specialized materials. The SPE class, which relies on amplitude detection ($|h|$), can operate coherently or incoherently depending on whether phase information is preserved. However, neither existing AO nor SPE implementations simultaneously provide phase preservation, tunable amplitude nonlinearity, stable power compensation, and residual connectivity. This motivates the design of the C3 unit, which integrates coherent SPE-based amplitude nonlinearity with per-layer power compensation and native optical residuals, directly embodying the theoretical conditions for scalable coherent PNNs.

The C3 unit mainly consists of a multi-waveguide resonator integrated with a detection-feedback (DF) branch (Fig. 3a). The input field $s^l$ from the $l$-th layer is split such that the majority of the signal $s_{in,1}$ enters the resonator and preserves its phase information, while the remaining portion $s_{DF}$ is directed to a photodetector whose current drives a feedback circuit. This loop dynamically tunes the resonator's loss and resonance, thereby generating a controlled amplitude nonlinearity without disturbing the phase relations. Simultaneously, one or more local oscillator (LO) fields ($s_{in,i}, i \geq 2$) are injected into the resonator through additional waveguides. These auxiliary ports provide coherent power compensation, establishing a non-zero and tunable power fixed point ($q^* > 0$). In this role, the LO injection acts as the hardware counterpart of the theoretical bias term ($\sigma_b$). A compact dynamic model (Supplementary Section 4) based on time-domain coupled-mode theory (TCMT)[42] expresses the activation as:

$$s^{l+1} = \sum_{j=1}^{m} S_{1j}\left(\omega; I_{fb}(|s^l|^2)\right) s_{in,j} \qquad (2)$$

where the complex coefficients $S_{1j}$ encode both the nonlinear feedback response and the LO-based compensation. Importantly, LO injection allows $|s^{l+1}/s^l| > 1$, overcoming the intrinsic attenuation that limits multi-layer optical cascading and providing the hardware basis for stable power propagation.

The multi-waveguide geometry naturally supports optical residual connections (Fig. 3b), enabling information flow across nonadjacent layers (Supplementary Section5). In practice, the unit operates in two configurable modes: Bias-Adjusted Injection (BAI), which sets and stabilizes the power fixed point, and Residual Injection (RI), which preserves complex-field correlation $c^*<1$. These two functions, per-layer stability and inter-layer routing, directly embody the theoretical conditions for scalable coherent networks.

**Unit fabrication and Nonlinear Activation.** Fig. 4a illustrates a two-port C3 unit (C3-$m$2) fabricated on a SOI platform using standard foundry processes (Methods). An on-chip Mach–Zehnder modulator (MZM) and phase shifter (PS) prepare the signal and LO inputs with tunable amplitude ratios and phases. Carrier-injection feedback produces an efficient, controllable optical nonlinearity without requiring external electrical amplification. The resonator's tuning characteristics, calibrated experimentally (Supplementary Sections 6), were incorporated into the activation model and implemented as a differentiable module compatible with common deep-learning frameworks (Supplementary Sections 7).

Operating the resonator under forward bias while tuning the LO injection, the feedback strength, and the optical frequency enables independent control of nonlinear shaping and loss compensation, consistent with the activation behavior described in Eq. (2). This removes the conventional trade-off between transmittance and nonlinearity in coherent platforms. The resulting activation functions span a wide range of surfaces (Fig. 4b and Fig. S6). The output exhibits strong, reconfigurable amplitude nonlinearity with a well-defined operating point (Fig. 4c). More details on data processing are shown in Supplementary Section 8.

To validate computational capability, we constructed a CoP-ResNet (Fig.4d)

using C3 units as the nonlinear layers while modeling all linear layers as lossy. The network exhibits stable convergence over deep layers (Fig. 4e), confirming that the C3 unit realizes the nonzero fixed points required for deep, expressive coherent photonic networks.

**Scalable PNNs Training.** We evaluate the scalability of C3 under realistic photonic constraints in two stages. We first use a width-constrained 5-class spiral benchmark to probe parameter efficiency and robustness to optical-power variations (Fig. 5). We then scale to the higher-complexity 1623-class Omniglot task to validate the role of our photonic residual connectivity (Fig. 6).

The spiral-task architecture is shown in Fig. 5a. We compare a coherent linear baseline with several activation functions (Supplementary Section 9) under varying layer widths. As shown in Fig. 5b, introducing a nonlinear stage yields a clear accuracy gain over the coherent linear baseline across all tested widths, demonstrating that linear propagation alone is insufficient for this task. In the width-constrained regime, Coherent C3 with LO injection consistently achieves the highest accuracy, indicating improved parameter utilization when coherent nonlinear expressivity is coupled with energy stabilization. To explain these trends, we examine the layer-wise evolution of optical power and effective nonlinearity strength (Fig. 5c). Without LO injection, attenuation causes power decay and progressively weakens the nonlinear response, making the coherent nonlinearity increasingly contractive with depth. In contrast, LO injection stabilizes both power and nonlinearity across layers, sustaining nonlinear expressivity without requiring additional optical gain. This stabilization translates into strong robustness to input-power variations (Fig. 5d), where Coherent C3 with LO injection maintains high accuracy over a broad power range, whereas the LO-free mode degrades rapidly as reduced signal amplitude deactivates nonlinear units. Notably, although the LO-free coherent mode weakens with depth, it can still exceed the incoherent variant at sufficiently large widths (e.g., $N = 32$), suggesting that increased layer width partially compensates for reduced deep-layer nonlinearity.

For the high-complexity Omniglot task, we implemented a SPE MLP-Mixer[43] architecture (Fig. 6a; Supplementary Section 10). Without residual connections, all nonlinearities—including electronic baselines—yield accuracies below 16%. Introducing C3-enabled optical residuals raises accuracy to 77.92%, surpassing both optical networks without residuals and electronic residual networks using complex-valued GELU. Consistent with the spiral results, the coherent C3 nonlinearity provides the necessary expressive capacity, while residual connectivity becomes essential for maintaining trainability at depth under loss. These experiments jointly demonstrate that C3 integrates coherent nonlinear expressivity with practical power stability, enabling scalable training of deep photonic neural networks in realistic settings.

**Discussion**

In this work, we establish a theoretical and experimental framework for depth-scalable PNNs grounded in statistical field theory. By analyzing the propagation of coherent optical fields in deep architectures, we demonstrate that collective system dynamics reduce to two macroscopic order parameters, revealing that scalable coherent photonic computation necessitates coherent amplitude nonlinearities to maintain an expressive dynamical regime. Our analysis further identifies a fundamental physical constraint: the contractive nature of lossy linear transformations. These lossy operators progressively compress the signal manifold, driving dynamics toward the ordered phase where expressive capacity vanishes. Compensating for optical power loss alone cannot reverse this collapse, as uniform amplification merely rescales the signal without recovering the geometric diversity of the high-dimensional state space. To counteract this contraction, we introduce residual connectivity as a correlation-preserving mechanism—naturally realized in coherent photonic systems through optical interference without electronic regeneration. Guided by these principles, we design and experimentally validate the C3 unit—a photonic module integrating coherent activation, loss compensation, and residual

connectivity—thereby demonstrating a viable pathway toward CoP-DNNs.

More broadly, our results establish a general physical framework for evaluating the scalability of optical computing architectures beyond the specific implementation. The analysis clarifies a cascaded design principle: the choice of information domain (real- versus complex-valued) governs the admissible class of nonlinearities (incoherent versus coherent amplitude operations), while the contraction mapping necessitates correlation-preserving mechanisms. These principles yield explicit criteria for evaluating and designing scalable optical systems, as summarized in Table 1. These results contribute to an emerging physics of scalable optical computation by redirecting attention from component-level performance metrics to fundamental physical constraints. Together, they point toward a transition from a photonics-as-accelerator paradigm to photonics as an analog computing system, enabling optical implementations of state-of-the-art large-scale AI models.

**Table 1 |** Design principles for scalable photonic computing architectures. The information domain determines the admissible nonlinearity class; lossy operation necessitates correlation-preserving mechanisms.

| Architecture Type | Nonlinear Design | | Structure Design | | Scalability Limitation |
|---|---|---|---|---|---|
| | Information Domain | Nonlinearity Type | Contraction Mapping | Correlation-Preserving Mechanism | |
| All-electronic (Classical) | Real-valued | — | No | — | Latency, bandwidth, energy |
| Hybrid electro-optic | Real-valued | Incoherent amplitude-dependent | Yes | Yes | Noise, latency, O-E-O overhead |
| All-optical (Incoherent) | Real-valued | Incoherent amplitude-dependent | Yes | Yes | Noise |
| All-optical (Coherent) | Complex-valued | Coherent amplitude-dependent | Yes | Yes (Our work: Photonic Residual) | Noise |

## Methods

**Device fabrication.** The C3-*m*2 units were fabricated in a standard 90-nm silicon photonics foundry using a silicon-on-insulator (SOI) wafer with a 220-nm-thick silicon-core layer and a 2-µm-thick buried oxide (BOX). Waveguides were patterned through a three-mask lithography and dry-etch sequence defining strip and rib geometries. The microring resonator modulators (MRM) incorporated PN junctions formed by P and N ion implantations, with P++ and N++ regions used for ohmic contacts. A vertical Ge/Si photodetector was monolithically integrated beside the MRM, realized through selective epitaxial growth of ~4 µm germanium on the silicon region. The silicon base was n-type doped, while the overlying Ge layer received p-type implantation, forming a vertical p-Ge/n-Si junction optimized for efficient carrier collection. Finally, silicon contacts, metallization, and pad passivation were implemented for electrical connectivity.

## Data availability

All the data that supports the findings of this study are available from the figures and from the corresponding authors upon reasonable request.

## Code availability

All the data that supports the findings of this study are available from the figures and from the corresponding authors upon reasonable request.

## Acknowledgements

This work is funded by National Key Research and Development Program of China (2021YFB2801700, 2021YFB2801702), National Natural Science Foundation of China (62341508, 62375240, 62175214), Leading Innovative and Entrepreneur Team Introduction Program of Zhejiang (2021R01001), Zhejiang Provincial Natural Science Foundation of China (LDT23F04012F05, LDT23F04015F05, LDT23F04014F01), Fundamental Research Funds for the Central Universities (226202400171), and Startup Foundation for Hundred-Talent Program of Zhejiang University.

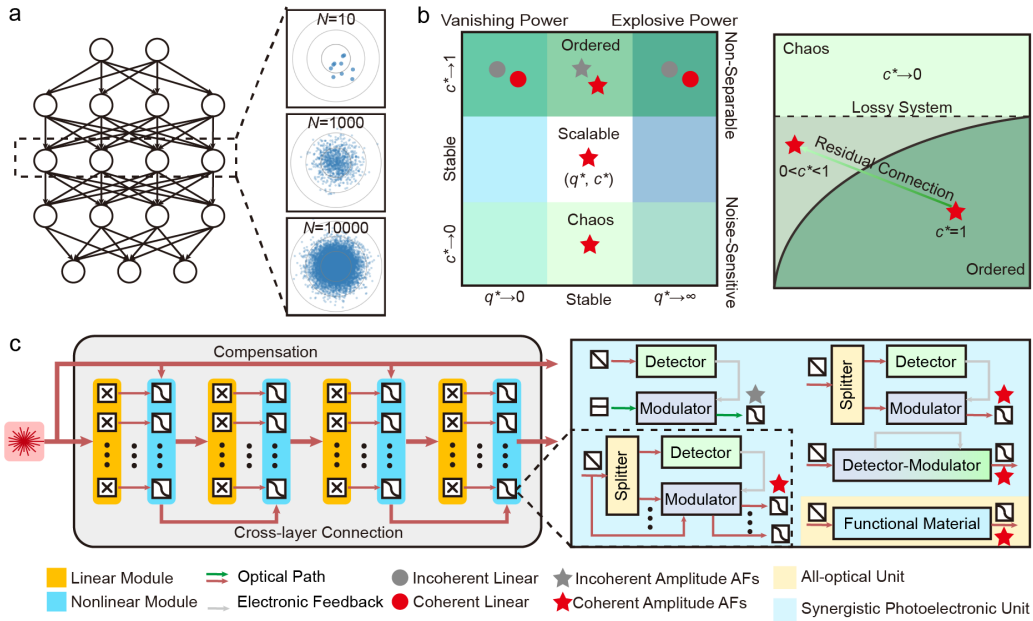

**Fig. 1 | From microscopic randomness to scalable coherent photonic computation. a**, Statistical emergence of macroscopic order. As the network width increases, the pre-activation fields converge towards a well-defined statistical distributions. **b,** Phase space and dynamical regimes. The state space defined by the order parameters ($q^*$, $c^*$) reveals distinct regimes governed by the type of nonlinearity. **c,** Implementation in scalable photonics. Schematic of different nonlinear activation function (AF) units and a scalable, coherent photonic neural network architecture.

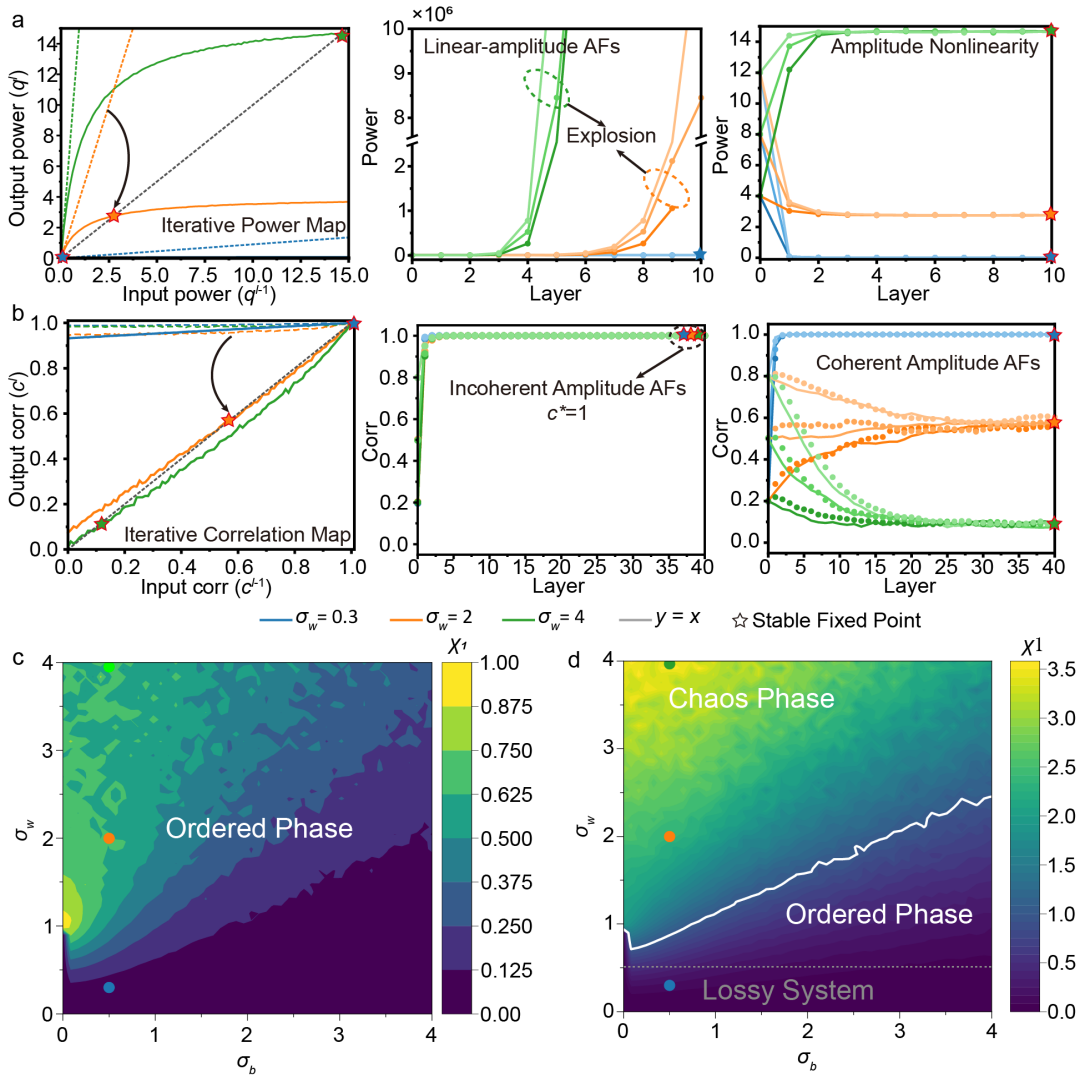

**Fig. 2 | Mean-field dynamics of coherent photonic networks. a, Dynamics of the power order parameter $q$.** The theoretical $\mathcal{Q}$-map (left panel, Eq.(1)) shows that linear activation functions, whether coherent or incoherent ($\sigma_b = 0$), cannot sustain a non-zero fixed point. In contrast, saturating nonlinearities (e.g., tanh) produce a stable fixed point $q^*$. Numerical simulations of wide networks (dots, $N_l = 1000$, Eq. (S1)) confirm the theoretical predictions (solid lines). **b,** Dynamics of the correlation order parameter $c$. The $\mathcal{C}$-map (left panel) demonstrates that incoherent nonlinearities drive correlations to unity ($c^* = 1$). Only coherent amplitude nonlinearities yield a stable, non-trivial fixed point ($0 < c^* < 1$), as validated by simulations. **c,** Phase diagrams. The stability of the correlation fixed point is governed by $\chi_1$, the slope of $\mathcal{C}$-map at $c = 1$. The critical line $\chi_1 = 1$ (white) separates the chaotic phase ($\chi_1 > 1$, where $c^* < 1$) from the ordered phase ($\chi_1 < 1$, where $c^* = 1$).

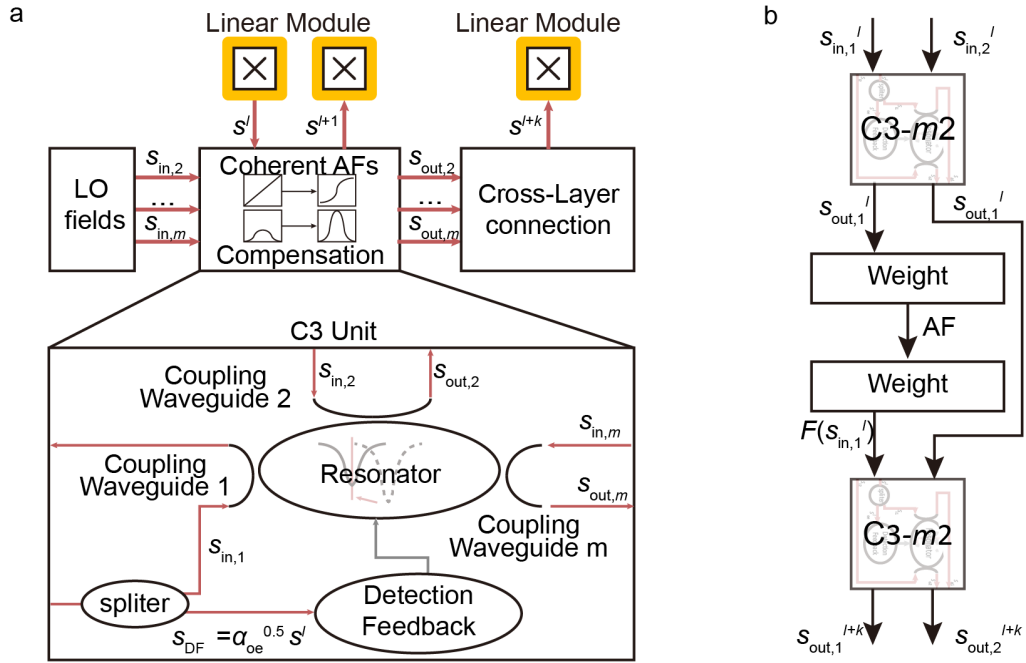

**Fig. 3 | Design of the C3 unit. a,** Schematic of the C3 unit, the operation workflow comprises three stages: an optical splitter splits the input field $s^l$ from the $l$-th layer into a detection arm ($s_{DF}$) feeding into the DF branch, and a transmission arm ($s_{in,1}$) injected into the resonator; the DF branch converts $s_{DF}$ to a photocurrent $I_p$, generating a closed-loop feedback signal $I_{fb}$ to tune the resonator's optical parameters; the resonator combine $s_{in,1}$ with ($m-1$) local oscillator (LO) fields through $m$ coupling waveguides, producing $m$-path coherent nonlinear optical outputs. **b,** Photonic residual block constructed with two C3 units.

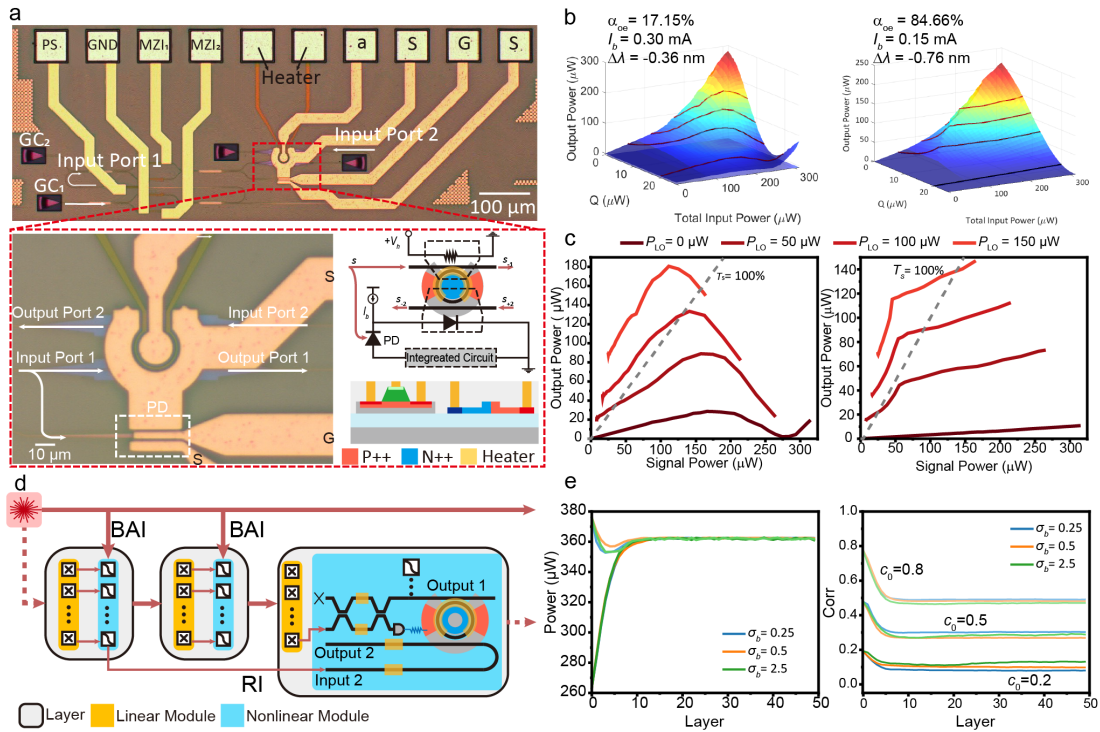

**Fig. 4 | Device realization and functional characterization of the C3 unit. a,** Optical micrograph of a fabricated C3-*m*2 unit, where *m* denotes the number of coupling waveguides. **b,** Measured nonlinear response surfaces showing output power versus total input power (x-axis) and signal-power splitting ratio governed by the modulation power of the Mach–Zehnder modulator (y-axis). Left and right panels correspond to bias currents $I_b$ = 0.3 mA and 0.15 mA, with photoelectric-conversion splitting ratios $α_{oe}$ = 17.15% and $α_{oe}$ = 82.8%, respectively. **c,** Corresponding nonlinear activation functions extracted from **b** for different local-oscillator injection powers. **d,** Architecture of a C3-based optical residual block. **e,** Simulated power and correlation dynamics of a coherent photonic residual network incorporating C3 units ($N_l$ = 1000, $σ_w$ = 0.5, $α_{oe}$ = 0.85, $I_b$ = 0.3 mA, $λ$=1553.5 nm, $|s_{LO}|$: $μ$ = 0.01, $σ_b$ ∈ {0.25, 0.5, 2.5}).

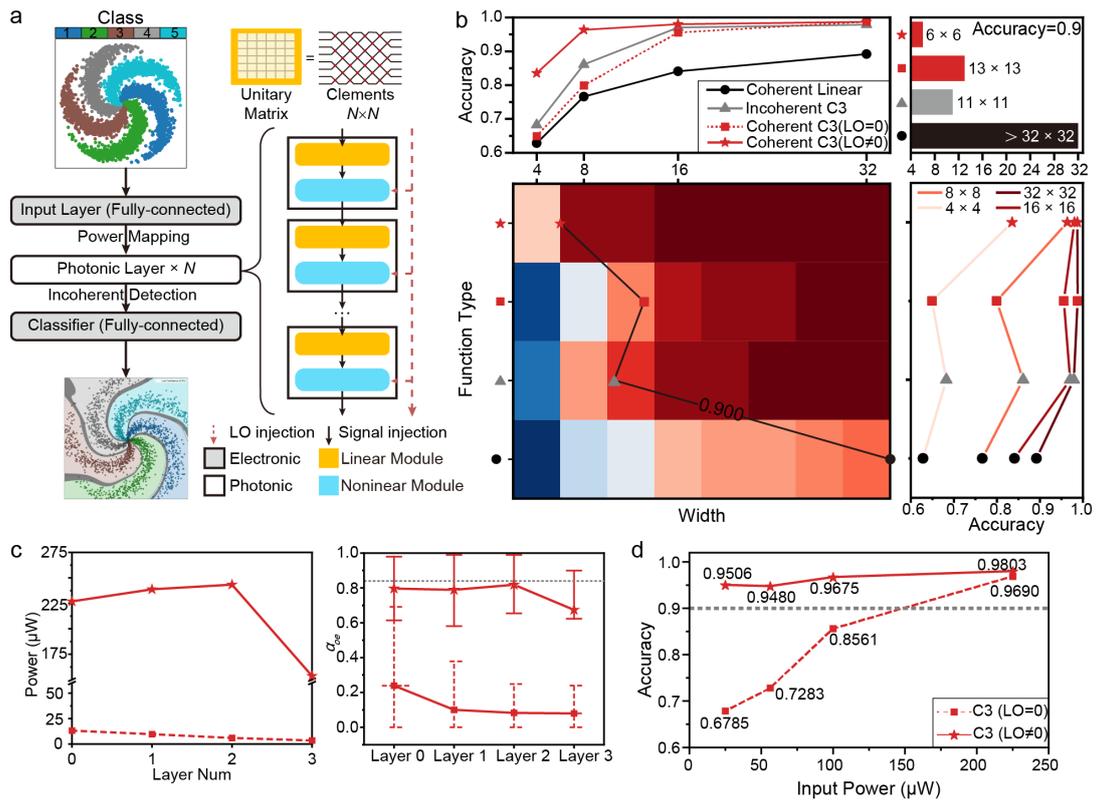

**Fig. 5 | Performance and robustness of C3-based photonic networks. a,** Architecture of the photonic network used for the five-class spiral task, comprising an electronic encoder, cascaded photonic layers, and an electronic classifier. **b,** Accuracy across activation functions and layer width at input power = 225 µW, showing the advantage of of C3 with local-oscillator (LO) injection in width-constrained settings. **c,** Layer-averaged optical power and C3 photoelectric activation ratio ($α_{oe}$) versus depth. Without LO injection (LO = 0), both power and $α_{oe}$ declines, indicating nonlinear deactivation. **d,** Classification accuracy versus input power, illustrating the power resilience of C3 (LO $\neq$ 0).

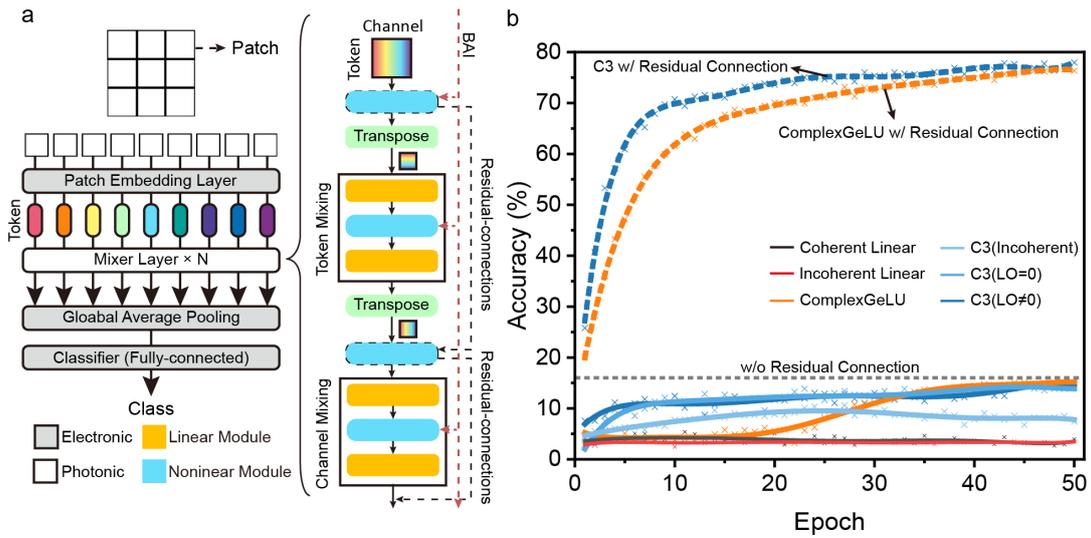

**Fig. 6 | SPE MLP-Mixer with C3 nonlinearities for Omniglot recognition. a,** Synergistic photoelectronic (SPE) MLP-Mixer architecture incorporating C3-based optical residual pathways. **b**, Test accuracy on the 1,623-class Omniglot dataset for different activation functions, showing the substantial performance gain enabled by C3-based optical residual connections.